\documentclass[12pt,preprint]{elsarticle}
\usepackage{fullpage}
\usepackage{natbib}
\usepackage{hyperref} 
\usepackage{graphicx}
\usepackage{amsfonts}
\usepackage{amssymb}
\usepackage{amsmath}
\usepackage{amsthm}

\newcommand{\N}{\mathbb N}

\newcommand{\G}{\mathcal G}
\newcommand{\Sgen}{\mathcal S}

\newtheorem{thm}{Theorem}
\newtheorem{defn}[thm]{Definition}
\newtheorem{prop}[thm]{Proposition}

\newtheorem{eg}[thm]{Example}

\newcommand{\etal}{{\it et al.}}

\bibpunct{(}{)}{;}{a}{}{,}

\title{Maximum likelihood estimates of pairwise rearrangement distances}

\author[crm]{Stuart Serdoz}
\ead{16115907@student.westernsydney.edu.au}
\author[crm,aiu]{Attila Egri-Nagy}
\ead{egri-nagy@aiu.ac.jp}
\author[tas]{Jeremy Sumner}
\ead{jeremy.sumner@utas.edu.au}
\author[tas]{Barbara R. Holland}
\ead{barbara.holland@utas.edu.au}
\author[tas]{Peter D. Jarvis}
\ead{peter.jarvis@utas.edu.au}
\author[babs,eerc]{Mark M. Tanaka}
\ead{m.tanaka@unsw.edu.au}
\author[crm]{Andrew R. Francis\corref{cor1}}
\ead{a.francis@westernsydney.edu.au}

\address[crm]{Centre for Research in Mathematics, Western Sydney University, Australia}
\address[aiu]{Current address: Akita International University, Japan}
\address[tas]{School of Physical Sciences, University of Tasmania, Australia}
\address[babs]{School of Biotechnology and Biomolecular Sciences, University of New South Wales, Australia}
\address[eerc]{Evolution \& Ecology Research Centre, University of New South Wales, Australia}

\cortext[cor1]{Corresponding author}

\begin{document}

\begin{abstract}
Accurate estimation of evolutionary distances between taxa is important for many phylogenetic reconstruction methods. Distances can be estimated using a range of different evolutionary models, from single nucleotide polymorphisms to large-scale genome rearrangements.
Corresponding corrections for genome rearrangement distances fall into 3 categories: Empirical computational studies, Bayesian/MCMC approaches, and combinatorial approaches.
Here, we introduce a maximum likelihood estimator for the inversion distance between a pair of genomes, using a group-theoretic approach to modelling inversions introduced recently. This MLE functions as a corrected distance: in particular, we show that because of the way sequences of inversions interact with each other, it is quite possible for minimal distance and MLE distance to differently order the distances of two genomes from a third.  
The second aspect tackles the problem of accounting for the symmetries of circular arrangements. While, generally, a frame of reference is locked, and all computation made accordingly, this work incorporates the action of the dihedral group so that distance estimates are free from any \emph{a priori} frame of reference. The philosophy of accounting for symmetries can be applied to any existing correction method, for which examples are offered.
\end{abstract}

\maketitle

\vspace{.5in}

\section{Introduction}
Estimates of evolutionary distance between pairs of taxa are key ingredients for reconstructing phylogenies, but are difficult to obtain reliably \citep{felsenstein2004inferring,gascuel2005mathematics}.  This is especially true for evolutionary models in which events can interact with each other in a way that affects inference. One estimate of distance between two genomes is  the \emph{minimal} distance which is model-specific and represents an assumption of parsimony in evolutionary paths through genome space (see \citet{fertin2009combinatorics} for a discussion of rearrangement models in this context). In fact, for most models, there are infinitely many possible evolutionary paths between any two genomes, and the minimal distance is simply the length of one of the shortest of these; by definition the minimal distance can only underestimate the true number of evolutionary events. 

The problems with using a minimal distance are well documented, especially when time periods are long and the space of obtainable genomes becomes saturated. Given enough time, all evolutionary endpoints become equally likely, and any signal of actual evolutionary time is lost. In some models, metrics have been developed to account for multiple changes; the most well-known perhaps being the Jukes-Cantor correction for models of single nucleotide substitution~\citep{jukes1969evolution}.  This method requires all events to be \textit{independent} (a common assumption with nucleotide substitution), but such independence  does not hold for most genome rearrangement models (such as inversion) and so alternative approaches are needed.  

Given pairwise distances obtained from a phylogenetic tree, \citet{buneman1971recovery} demonstrated that the recovered tree is unique, a fact which also follows from the 4-point condition~\citep{buneman1974note}. Furthermore, \citet{warnow1996some} and \citet{atteson1999performance} suggest that if the true evolutionary distance inference is sufficiently accurate, even polynomial time reconstruction algorithms, such as Neighbor Joining \citep{saitou1987neighbor}, will return the correct phylogeny. Recent work by \citet{gascuel2015stochastic} places the results of Atteson~\etal~ in a statistical framework.

Some studies attempt to find a relationship between true distance and minimal distance (or some other available measure such as breakpoint distance), and use this to produce an estimate of true distance as a function of minimal distance. For instance,~\citet{wang2001estimating} introduced an estimator of true evolutionary distance called \textit{IEBP} (inverting the expected breakpoint distance). The method operates under the generalised Nadeau-Taylor model \citep{nadeau1984lengths} and provides a robust polynomial time algorithm to estimate true evolutionary distance. Similarly, the \textit{EDE} (empirically derived estimator) of \citet{moret2001new} samples the relationship between inversion distance and true evolutionary distance before providing a fit. Applications of IEBP and EDE can be seen in \citet{li2002genome}. 

While a useful correction, such estimates are based on just one factor -- the minimal distance --  and can't account for underlying structure of the genome space (in our framework, the Cayley graph of the group).  The key point being that not all elements of equal minimal distance are equally likely.

As an optimal estimate of true distance, we would like (very loosely) some sort of \textit{expected} distance -- a function of final arrangement -- constructed as a weighted average of evolutionary paths, pushing the problem into the intersection of combinatorics and statistics. In this vein,~\citet{eriksen2002approximating} offered an approximation of the {expected} number of inversions to have occurred given $n$ breakpoints. This was followed by a method of estimating the expected inversion distance by looking at the expected transposition distance \citep{eriksen2004estimating}, and generalizations such as~\citet{eriksen2005expected} and \citet{dalevi2008expected}. 

Given the sizes of the spaces involved, MCMC and Bayesian methods play an important role. \citet{york2002bayesian} uses a Bayesian framework to estimate true distances for inversions.  
On the MCMC front,~\citet{miklos2003mcmc} introduced a time continuous stochastic approach to genome rearrangements (modelled as a Poisson process), allowing reliable estimates of true distances. The key aim being to describe the posterior distribution of true evolutionary distance given two arrangements. There have been several generalizations to these methods: \citet{durrett2004bayesian} includes translocations as well as inversions; \citet{Larget2005} describe a  Bayesian method for phylogeny inference and offer a comparison between their approach and a parsimony approach; and~\citet{miklos2009efficient} provide a method to estimate the \emph{number} of minimal walks.

This paper describes a novel \emph{maximum likelihood} approach to corrected rearrangement distances. We focus on models of genome rearrangement involving invertible operations, such as inversion and translocation, which can be described in group-theoretic terms, using the framework introduced in~\citet{egri2014group} and~\citet{francis2014algebraic}. This algebraic framework treats genomes as the images of the actions of elements of a finite reflection group, and allows us to treat the genome as not fixed in space, but free to rotate in Euclidean space. Each genome is then considered to be a coset in the quotient of the main reflection group by the dihedral group.  

The next section describes the general group-theoretic models of chromosome rearrangements on which this paper is based. The third section introduces the likelihood function under our model, and gives some basic examples of what these functions look like. Next, we compare the minimal distance to the MLE and give an example of how the resulting phylogenetic inference can give different results.  We then consider properties of group elements that may characterise the likelihood function and hence the MLE of distance. The penultimate section describes what is required to account for dihedral symmetry, and illustrates the approach with some example phylogenies. We end with a discussion of some of the issues involved in using the MLE.

\section{Group theoretic models of rearrangement}\label{s:group}
In this section we describe {group-theoretic} models of genome rearrangement, following the development in~\citet{egri2014group}. Such models allow events that change the underlying sequence in a reversible way, including for example inversion and translocation but not insertion or excision.  The invertible rearrangements defined by the model then generate a \emph{group}, and there is a one-to-one correspondence between the set of possible genome arrangements and the set of elements of this group.

This correspondence in practice requires two additional assumptions.  First, we choose one genome as the reference genome, that will correspond to the group identity element.  This is arbitrary, and is discussed in more detail below.  Second, we assume there is no rotation of the genome in 3-dimensional space.  We think of this as fixing a ``frame of reference'' for all genomes.  This assumption is removed for calculating MLEs of evolutionary distances in ways described below, by taking a quotient by the dihedral group.

The genome space is then realized as a graph with genomes as vertices and allowable evolutionary events defining edges between them. This corresponds to a graph based on the group, called the \emph{Cayley graph}, whose vertices are group elements and edges represent multiplication by the group generators. Thus the Cayley graph can be thought of as a map of the genome space, with vertices the possible genomes (group elements) and edges the possible rearrangement events (generators of the group)~\citep{clark2016bacterial}. The Cayley graph depends on both the group $\G$ and the generating set $\Sgen$. 

Given a choice of one arrangement as the reference genome $G_0$, every other genome arrangement can be obtained from $G_0$ by a sequence of rearrangements.  Because each allowable rearrangement event defines a generator of the group, this sequence of rearrangements is a product of group generators, and therefore corresponds to a group element itself.  Thus the reference genome $G_0$ corresponds to the identity element $e$ of the group $\G$, and each other possible genome corresponds to a unique group element (remembering that for now we assume a fixed frame of reference).  Note that there may be many sequences of events giving rise to the same genome, and these correspond to different walks through the Cayley graph. 

A brief note on the language of paths and walks.  In graph theory a \emph{walk} through a graph is an alternating sequence of vertices and edges beginning at one vertex and ending at another.  This may or may not involve traversing the same edge or vertex multiple times. A \emph{path} on the other hand is a walk in which no vertices or edges are visited more than once.  To avoid confusion we will use ``walk'' in the context of the Cayley graph, but it is worth noting that \emph{minimal} walks between two group elements on the Cayley graph are all paths, in this sense.  It is common, however, outside of graph theory, to use the expression ``evolutionary path'' between two organisms without the implication that no genome has been visited more than once (allowing, for instance, homoplasy or convergent evolution), and we will also use ``path'' in that context, where clear.

Returning to walks and distances on the Cayley graph, observe that the choice of reference genome is not important.  For any two genomes $G_1$ and $G_2$ with corresponding group elements $g_1$ and $g_2$, there is a unique group element (namely $g_1^{-1}g_2$ when acting on the right) that transforms $G_1$ into $G_2$. 
As a result of the transitive group action \citep{babai1996automorphism}, the group element is independent of the choice of reference genome.  For instance if $G_1$ was chosen as the reference genome then the walk from $G_1$ to $G_2$ would still correspond to the group element $g_1^{-1}g_2$ (which in this case would be simply $g_2$, since here $g_1=e$). 

With this correspondence between the genome space and the Cayley graph, the \textit{minimal distance} (denoted $d_{min}$) on the genome space, the ``word metric'' on the group \citep{lyndon2015combinatorial}, and the path metric on the Cayley graph all coincide.  

An evolutionary history between two genomes is a random walk on the genome space using allowable rearrangements, or equivalently, a random walk on the Cayley graph --- a well-studied topic~\citep{aldous2002reversible,lubotzky1995cayley,godsil2001algebraic}.  
Such a walk corresponds to a sequence of (right) multiplications of the group element at the starting point by the generators labelling the edges on the walk.  That is, an evolutionary path from $g_1$ to $g_2$ takes the form of the initial genome followed by a concatenation on the right of the applied events.  For example, a walk along the edges from $g_1$ beginning with $s_2$ and subsequently the sequence of generators $s_5, s_2, s_1, s_7$ corresponds to the equation in the group given by
\(
g_1 s_2 s_5 s_2 s_1 s_7 = g_2.
\)
This corresponds to a walk of length 5. Each such product of group generators from a walk between $g_1$ and $g_2$ represents the same group element, namely $g_1^{-1}g_2$.

The transitivity that we mentioned earlier means that walks from $g_1$ to $g_2$ are in correspondence with walks from the identity $e$ to $g_1^{-1}g_2$, and so it is sufficient to study walks and distances from the identity to a group element $g$.   

There are infinitely many walks to a group element, each giving a distinct word in the generators of the group (labels on edges of the Cayley graph), since walk length is unbounded.  A \emph{reduced} word is one that corresponds to a minimal length walk. Any word in the generators can be reduced to a minimal one using the group relations; these in turn correspond to loops in the Cayley graph. Generally, reduced words are not unique: there may be many ``parsimonious'' walks of minimal length. For further reading on the interaction between relations and words see \citet{lyndon2015combinatorial}. In what follows we will not be just interested in walks of minimal length, but in all walks between two genomes. 

In this set-up, given a random walk from $e$ to $g$, the minimal distance is the length of a geodesic (minimal) walk, while the true evolutionary distance is the length of the actual walk. The model of rearrangement we use as an example throughout this paper is the 2-inversion model studied by~\citet{egri2014group}, in which adjacent regions are swapped, and orientation is ignored.  However our general principles apply to any group-theoretic rearrangement system in which the generators are of order 2 (that is, the basic evolutionary events undo themselves if applied twice), and in particular any model of inversions.

\begin{figure}[ht]
\centering
\includegraphics[width=.4\textwidth]{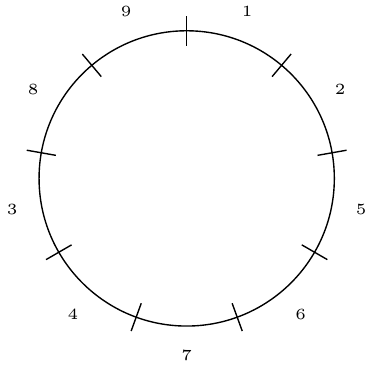}
\caption{Cycle notation tracks the movement of individual regions. The genome above is represented by $g = (3,7,5)(4,6)$.}\label{f:cycle.notation}
\end{figure}

When referencing specific circular genome arrangements we will use cycle notation in which the cycle $(\dots,a,b,c,\dots)$ means ``\dots, region $a$ is in position $b$, and region $b$ is in position $c$, \dots''.  For instance the permutation shown in Figure~\ref{f:cycle.notation} is represented by $g = (3,7,5)(4,6)$, which we read ``region 3 is in position 7, region 7 is in position 5, region 5 is in position 3; region 4 is in position 6 and region 6 is in position 4''.  Swapping regions 4 and 5 is done by multiplying on the right by the generator $(4,5)$, which gives the result $(3,7,5)(4,6)(4,5)=(3,7,4,6,5)$ (the reader may draw this to convince herself that this has the desired result).

\begin{center}
FIGURE 1 AROUND HERE.
\end{center}

\subsection{Genomes in three dimensional space}
Fixing the frame of reference makes for a clean translation of genome space to the Cayley graph, but in fact if two arrangements can be obtained from each other by rotation in three dimensional space, then biologically they are the same arrangement.  Mathematically, this means that if two circular genomes can be reached from each other by rotating the circle or reflecting it in an axis of symmetry then they are equivalent genomes.  These actions (rotation and reflection in an axis) generate the dihedral group $D_{n}$, which is a subgroup of the full group of rearrangements $\mathcal G$.  

The equivalence defined by such rotations and reflections amounts to taking a quotient of the group arising from the fixed-frame genome space by the dihedral group.  Two genomes being reachable from each other by such actions means that their corresponding group elements are in the same coset in this quotient.  In other words, each equivalence class of genomes under three-dimensional rotations corresponds to a coset in $\mathcal G/D_n$.  Put very bluntly, ``genomes are cosets''. Each element $g \in \mathcal{G}$ lies in a unique coset $gD_n = \{gd~|~d \in D_n\}$, so that elements within a coset represent the same arrangement. Instead of mapping the genomes to group elements; to account for symmetry the genomes are instead mapped to \emph{cosets} of $\G$ under $D_n$. 

In this light, a measure of distance between two genomes needs to be the distance between the \emph{cosets} corresponding to each genome.  This is implemented for minimal distances in~\citet{egri2014bacterial}.  The minimal distance is found by searching through all ordered pairs of coset representatives, finding the pair which minimises the minimal distance. 
Formally, for cosets $g_1D_n$ and $g_2D_n$, the minimal distance is given by
\[d_{min}(g_1D_n,g_2D_n) = \min\{~d(h_1,h_2) \mid h_1 \in g_1D_n; h_2 \in g_2D_n~\}.\]
In fact, thanks to group transitivity, we need not check all pairs: it suffices to compare the minimal distances from the elements of one coset to an arbitrary fixed representative of the other coset.  That is, while for $n$ regions the above expression would suggest $4n^2$ calculations, the transitivity of the Cayley graph reduces this to $2n$.

Finally, we note that while it may seem that bacteria have an internal frame of reference provided by regions such as the origin of replication \emph{ori}, incorporating this feature does not save computational effort or simplify the model.  If we wish to fix \emph{ori} as the reference point, then an inversion on an $n$-region genome that exchanges \emph{ori} with an adjacent region must be modelled as an $(n-1)$-cycle.  The group generators would then be $n-1$ adjacent transpositions, together with an $(n-1)$-cycle.  There would then be only two symmetries, obtained by reflecting the circular genome (not rotating), so the number of distinct arrangements would be $(n-1)!/2$, since there are $(n-1)!$ distinct ways to arrange the non-\emph{ori} regions around the rest of the genome.  In contrast the model described above allows \emph{ori} to move freely but accounts for symmetries later, meaning the generators of the group are $n$ adjacent transpositions (no longer-length cycles), but there are $2n$ symmetries.  The number of distinct arrangements is still $n!/2n=(n-1)!/2$.  The latter model is the one used in this paper, because of its capacity to exploit results in the theory of Coxeter groups.  

\section{Likelihood functions for group elements}\label{s:likelihood}
In this section we introduce an alternative to the minimal distance by establishing a maximum likelihood approach to estimating evolutionary distance. 
While genomes are regarded as cosets of group elements that are equivalent under the dihedral group action, we begin with considering group elements alone, before building the genome likelihood functions based on their respective cosets.

An important point with regard to group-based genome rearrangement models is that the group is generally \emph{non-abelian}, which means that operations do not commute.  This arises directly from the biological model: the effect of two successive inversions that overlap depends on the order in which they are done.
When it comes to random walks, the key impact is that with abelian groups all endpoints of a walk of a given length are equally likely.  This is not the case for non-abelian groups such as those generated by inversion models, as can be seen in the following example. 

\begin{eg}
Consider the group $\G$ generated by linear (as opposed to circular) 2-inversions over 9 regions so that $\Sgen = \{s_i = (i,i+1)~|~i \in {1,\ldots,8} \}$. Both group elements $g_1 = (1,4,3,2)$ and $g_2 = (2,3)(4,5)(8,9)$ have minimal distance of three. While $g_1$ can only be realised by the sequence $s_1s_2s_3$; $g_2$ can be realised by $s_2s_4s_8,\, s_4s_2s_8,\, s_8s_4s_2,$ and more. This particular example relies on the fact that disjoint cycles commute.  
\end{eg} 

Let $g$ be a genome arrangement, with $n$ the number of distinct rearrangement events allowed by the model. Write $\alpha_i(g)$ for the number of walks from $e$ to $g$ of length $i\in\N$. Parameterise $\lambda=rT$ where $T$ is time and $r$ the number of rearrangements per unit time. Then the likelihood 
of $\lambda$ given the walk ends at $g$ is given by 
\begin{align}
	L(\lambda~|~g) 	&= Pr(g~|~\lambda) \nonumber \\
			&= \sum_{i \geq 0} Pr(g~|~i)~Pr(i~|~\lambda). \label{eqn:P(GgivenI)}
\end{align}
Assuming that time between events follows an exponential distribution, we have $Pr(i~|~\lambda) = e^{-\lambda}\lambda^i / i!$.  The assumption that walks of equal length are equally likely forces $Pr(g~|~i) = \alpha_i(g) / n^i$, and so

\begin{align}\label{eqn:likelihood}
	L(\lambda~|~g) 	&= \sum_{i \geq 0} \frac{e^{-\lambda} \lambda^i}{i!} \frac{\alpha_i(g)}{n^i}.
\end{align}
Maximising this function with respect to $\lambda$ gives a maximum likelihood estimate $\widehat\lambda$ of this parameter. In some special cases, closed-form expressions for $\alpha_i(g)$ may yield closed form likelihood functions (such as in Example~\ref{eg:s3}).  In other cases, $\widehat\lambda$ can be obtained numerically.

\begin{eg}\label{eg:s3}
A circular genome with only three regions, evolving under a model of inversions of adjacent pairs of regions, corresponds to the action of the symmetric group $S_3$ with circular generators $\{(1,2),(2,3),(3,1)\}$.  One can show that $\alpha_i(g)=3^{i-1}$ if $i$ and the permutation $g$ are both even or both odd, and if $i\ge d_{min}(g)\ge 1$, and it is zero for other cases with $i\ge 1$.  In the edge case $i=0$, $\alpha_0(g)=1$ for $g=()$, since there is one (empty) path of length 0 from the identity to itself, but is zero for other group elements.

In this simple example the likelihood functions for the group elements are 
\begin{align*}
	L( \lambda ~|~ ()) 	&= \frac{e^{-\lambda}}{3} \left[3 + \frac{\lambda^2}{2!} + \frac{\lambda^4}{4!} \ldots \right] 
						= \frac{e^{-\lambda}}{3} \left[2 + \cosh{\lambda}\right];  \\
	L( \lambda ~|~ (1,2)) 	&= \frac{e^{-\lambda}}{3} \left[\frac{\lambda^1}{1!} + \frac{\lambda^3}{3!} + \frac{\lambda^5}{5!} \ldots \right]  
							= \frac{e^{-\lambda}}{3} \sinh{\lambda}; \\
	L( \lambda ~|~ (1,2,3))	&= \frac{e^{-\lambda}}{3} \left[\frac{\lambda^2}{2!} + \frac{\lambda^4}{4!} + \frac{\lambda^6}{6!} \ldots \right]  
							= \frac{e^{-\lambda}}{3} \left[ \cosh{\lambda} - 1 \right].
\end{align*}
\end{eg}

The likelihood functions in Example~\ref{eg:s3} are monotonic in $\lambda$; something that is not true in general. For models with more regions, no closed form expressions are known, and hence the likelihood functions must be approximated by truncating the series.  

Unlike cases such as Example~\ref{eg:s3}, the walk-count function $\alpha_i(g)$ does not usually have a closed form.  It can, however, be computed using a simple recursive algorithm.  Suppose we want to count the number of walks of length $i$ that end at $g$.  Each such walk goes through an immediate neighbour of $g$, after having traversed a walk of length $i-1$.  Therefore the number of walks of length $i$ to $g$ is the sum of the numbers of walks of length $i-1$ to the immediate neighbours of $g$.  While there are some economies that can be made to this recursion (for instance we may know that some of the walk-counts are zero), it is still a computationally demanding algorithm, and only currently effective in practice for models of up to nine regions.

\section{Minimal distance and the MLE}\label{sec:min.and.mle}

In general, phylogenetic distance methods assume some relationship between distances and evolutionary time. That is, all methods presume that a larger distance implies a greater time since evolutionary divergence. 

One way to investigate this relationship is to compare the relative orderings placed by the metric on the set of pairs of genomes.  While distance based phylogeny reconstruction methods do not rely solely on the ordering of distances, they are sensitive to it. 

\begin{center}
FIGURE 2 AROUND HERE.
\end{center}

\begin{figure}[!ht]
\centering
\includegraphics[width=\textwidth]{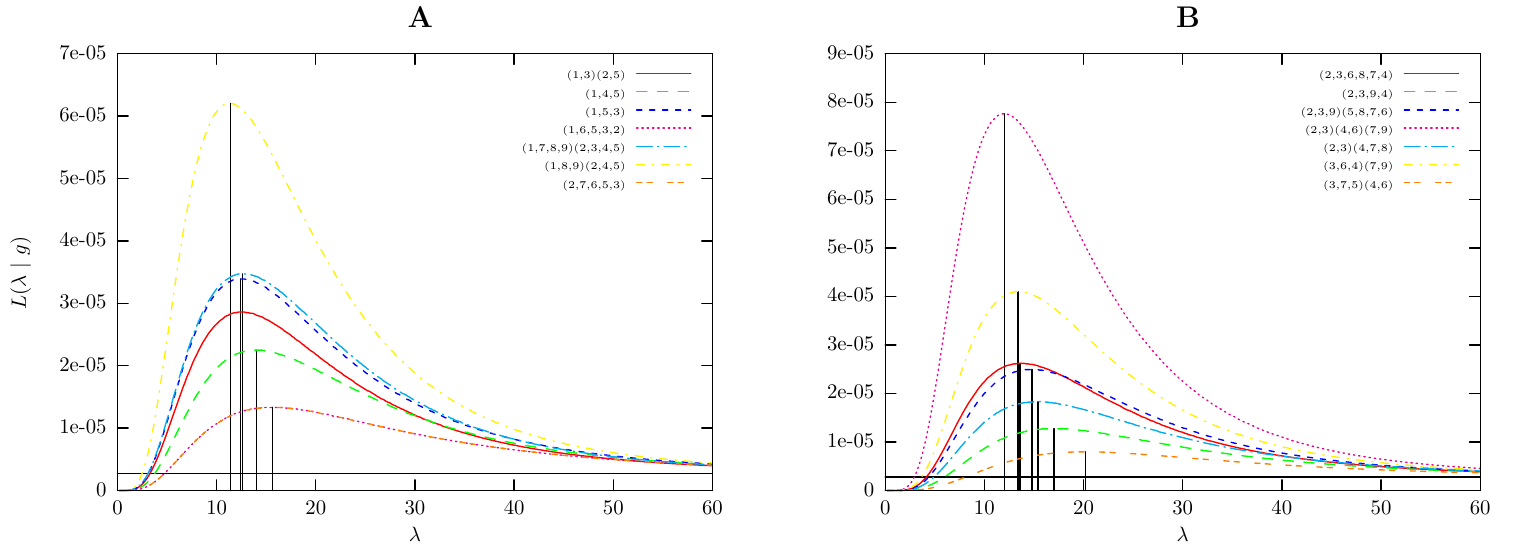}
\caption{Samples of elements of minimal distance six (A), and minimal distance seven (B).  The estimates, shown as vertical lines for each likelihood curve,  clearly overlap and in some cases the partial order reversal is stark: $d_{min}\left((2,7,6,5,3)\right) = 6$ and $d_{min}\left((2,3)(4,6)(7,9)\right) = 7$ but $\widehat\lambda_{(2,7,6,5,3)} \approx 16$ and $\widehat\lambda_{(2,3)(4,6)(7,9)} \approx 12$.}
\label{f:partial_orders}
\end{figure}

Figure~\ref{f:partial_orders} highlights examples where the partial order under the minimal distance and the partial order under the MLE differ ( i.e. $d_{min}(g_1) > d_{min}(g_2) \text{ but } \widehat\lambda_{g_1} < \widehat\lambda_{g_{2}}$). This reversal of partial order relations is (given our assumptions) a function of walk counts. These examples are not uncommon and highlight the problem with the minimal distance: it does little to characterise the MLE. It is not difficult to construct examples where this gives rise to differences in phylogenetic inference (see Figure~\ref{f:phylos}). We turn our attention to the question of what conditions on arrangements give rise to the same MLE.

\section{What group elements have the same MLE?}\label{sec:when.same.mle}

The previous section demonstrated that the structure of the genome space, represented by the Cayley graph, makes minimal distance a poor proxy  of evolutionary time.  A natural question arises: are there features of arrangements that can predict the MLE?  
One possibility for two arrangements to have the same MLE is when they have identical likelihood functions, and likelihood functions are determined by walk counts (Eq.~\eqref{eqn:likelihood}). 
In~\citet{clark2016bacterial}, it is shown that if two group elements are conjugate under the normalizer of the generating set, then their set of minimal length walks are not only the same size, but also order isomorphic. An extension of this given in Proposition~\ref{prop} provides a sufficient condition among arrangements to ensure equality of the MLE.

\begin{defn}[Normalizer]
Let $\mathcal{G}$ be a group, and $X$ a subset of $\mathcal G$. The \emph{normalizer} of $X$ in $\mathcal G$ is defined as
\[ N_{\mathcal G} ( X) = \left\{ g \in \mathcal{G} ~|~ g^{-1} X g = X \right\}.\]
\end{defn}

\begin{prop}\label{prop}
Let $\mathcal G$ be a group generated by $\mathcal S$. Write $\sim_{N}$ to mean conjugate under an element of $N_{\mathcal G}(\mathcal S)$. For $g_1, g_2 \in \mathcal{G}$ we have
\[g_1 \sim_N g_2 \implies L(\lambda ~|~ g_1) \equiv L(\lambda ~|~ g_2) \implies \widehat\lambda_{g_1} = \widehat\lambda_{g_2}, \]
where $\lambda_1$ and $\lambda_2$ represent the MLE for $g_1$ and $g_2$ respectively.
\end{prop}

The proof can be found in Appendix~1.  

While this condition may seem formal and abstract, in fact it corresponds to a very intuitive action on the generators.  Conjugacy by the normaliser of the generating set amounts to a relabeling of the generators in such a way that the relative positions of the generators is preserved.
For an example consider our base generating set of circular adjacent transpositions. The group elements $(1,3,5), (2,4,6), (3,5,7), \ldots$ are all conjugate under the normaliser; they have identical walk counts, and hence {if} the distribution across generators is uniform, they are represented by identical likelihood functions.

\begin{figure}[!ht]
\centering
\includegraphics[width=\textwidth]{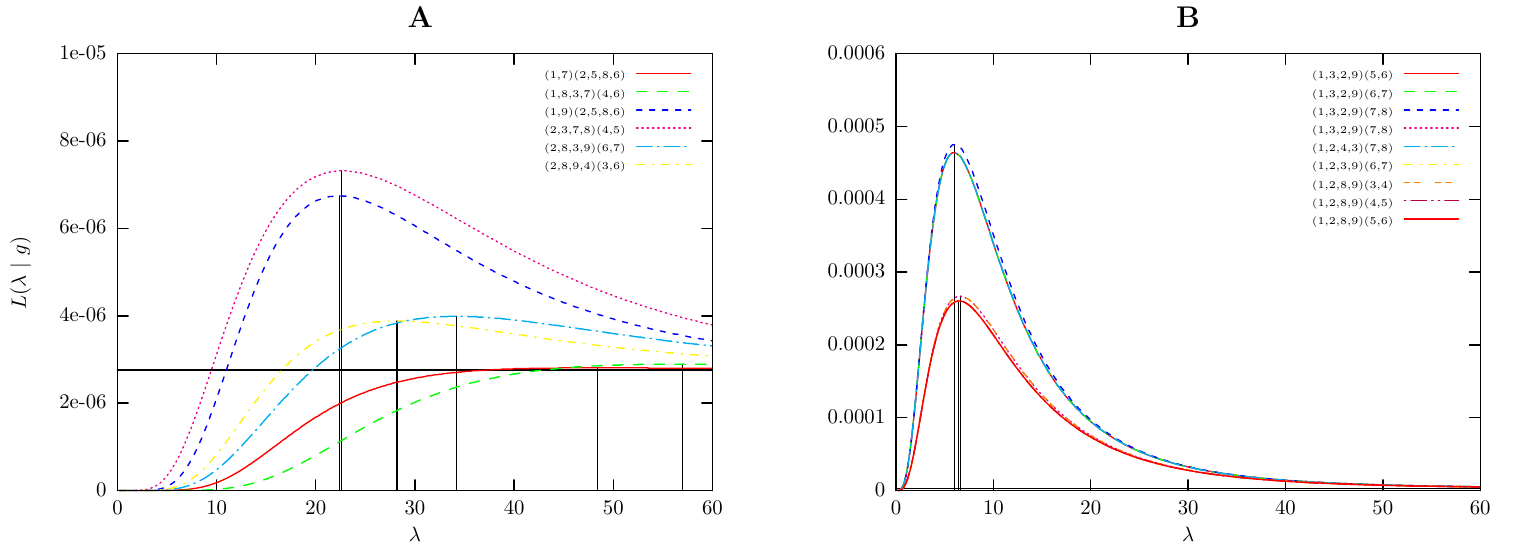}
\caption{(A)  Likelihood functions and MLEs for group elements of various $d_{min}$ from the conjugacy class $C$ with representative $(1,2,3,4)(5,6)$. (B) Likelihood functions and MLEs for group elements with identical minimal distance ($d_{min}(g)=4$) in $C$.  }\label{fig:conj}
\end{figure}

If elements are conjugate (but not by $N_{\G}(\Sgen)$) their likelihood functions are generally different (see Figure~\ref{fig:conj}) and are not guaranteed to give the same MLE.
In Example~\ref{eg:s3}, $N_{\G}(\Sgen)$ is the entire group, and in this case all elements of a conjugacy class will all also be conjugate under the normalizer. We now return to the calculation of likelihood functions for distances between two genomes, allowing movement in three dimensions.  

\begin{center}
FIGURE 3 AROUND HERE.
\end{center}

\section{MLE estimates of distance between genomes as cosets}\label{s:cosetMLE}

Given the likelihood functions defined above for particular group elements, we now introduce their application to cosets in order to address the issue of their symmetries in three dimensional space. 
As described above, to account for this, we consider each genome arrangement as a \emph{coset} of the group $\G$ under the dihedral group $D_n$. To construct a likelihood function for evolutionary distance between two cosets, in principle we need to consider all walks between the cosets.  Fortunately, we can use the results of the previous section to reduce the need to count walks between these $(2n)^2$ pairs of elements.  

Take two cosets $g_1D_n$ and $g_2D_n$ ($g_1,g_2\in G$), and let $\gamma$ be a walk of length $i$ between $g_1$ and $g_2$, so that $g_1\gamma=g_2$.  Then for any element of the dihedral group $d\in D_n$ we have $g_1\gamma d=g_2d$ and therefore 
\[
g_1d(d^{-1}\gamma d)=g_2d.
\]
Note that $g_1d\in g_1D_n$ and $g_2d\in g_2D_n$, so for each $d\in D_n$, $d^{-1}\gamma d$ defines another walk between the two cosets.

Since $D_n$ is a subgroup of the normaliser of the generating set, $N_\G(\mathcal S)$, the element $d^{-1}\gamma d$ remains a walk of length $i$, by the arguments in the proof of Proposition~\ref{prop}.
This means that to count walks of length $i$ between the two cosets, it is sufficient to choose a single representative of one of the cosets and consider walks of length $i$ to each of the $2n$ elements of the other coset.  In other words, instead of considering walks of length $i$ between $g_1D_n$ and $g_2D_n$, we may simply consider walks between $g_1$ and $g_2D_n$.

The other simplification that can be made is that as before, the transitivity of the Cayley graph under left multiplication means that instead of walks from $g_1$ to $g_2D_n$ we may instead count walks from the identity $e$ to $g_1^{-1}g_2D_n$.  The problem is reduced to considering walks of length $i$ from the identity to any coset $gD_n$. 

Let $g$ be one permutation representation of the genome, and $gD_n$ the corresponding coset.  Because walks to each coset element are independent from each other, the likelihood function splits into a sum across elements of the coset:
\begin{align*}
L(\lambda~|~X=gD_n) &= Pr(e\to gD_n\mid\lambda) \\
					&= \sum_{d \in D_n} Pr(gd~|~\lambda) \\
					&= \sum_{d \in D_n} \sum_{i\geq0} Pr(gd~|~i).Pr(i~|~\lambda)\\
					&= \sum_{d \in D_n} \sum_{i\geq0} \frac{\alpha_{gd}(i)}{n^i}\cdot \frac{e^{-\lambda}\lambda^i}{i!}.
\end{align*}
The last expression above has terms that are just the likelihoods for individual group elements, and so we have:

\begin{align}\label{eqn:cosetlikelihood}
L(\lambda~|~X=gD_n) &= \sum_{d \in D_n} L(\lambda\mid X=gd).
\end{align}
That is, the likelihood for a genome distance, allowing for three-dimensional rotations, is the sum of the individual likelihood functions for each of the group elements in the coset.

\section{Minimal distances and MLEs of distances between genomes.}\label{s:min.vs.mle}

We have now described several ways to define a distance between genomes under a group-theoretic model of rearrangement (such as using inversions).  Given one reference genome and another given by a group element $g$ (describing the permutation of the regions), these are: 
\begin{enumerate}
	\item The minimal distance to $g$,
	\item The minimal distance to the coset containing $g$ (the method developed in~\cite{egri2014group} to account for the genome in three-dimensional space),
	\item A maximum likelihood estimate for the distance to $g$, and 
	\item A maximum likelihood estimate for the distance to the coset containing $g$.
\end{enumerate}
The latter two have been introduced in this paper, and we have described the way the MLE for the distance to a group element (3) can provide more information than use of the minimal distance alone (1), in the sense that the ordering of elements by minimal distance is often not preserved when taking MLEs (Figure~\ref{f:partial_orders}).  

The same clash between minimal distance and MLE of distance arises when allowing the genome to rotate in three dimensions (the coset approach). 

For instance, consider the group element $(1,2)$ with minimal distance 1 from the identity.  If we rotate the arrangement on nine regions once, we obtain the group element $(1,3,4,5,6,7,8,9)$ (region 1 is in position 3, region 3 is in position 4 etc), which has minimal distance 7. Another rotation gives $(1,4,6,8)(2,3,5,7,9)$ with minimal distance 13.  In other words, a single coset can contain elements of very different minimal lengths.  This is a reminder that fixing the frame of reference and calculating the minimal distance between two genomes on the basis of just one frame is likely to result in large errors.

\begin{figure}[!ht]
\centering
\includegraphics[width=\textwidth]{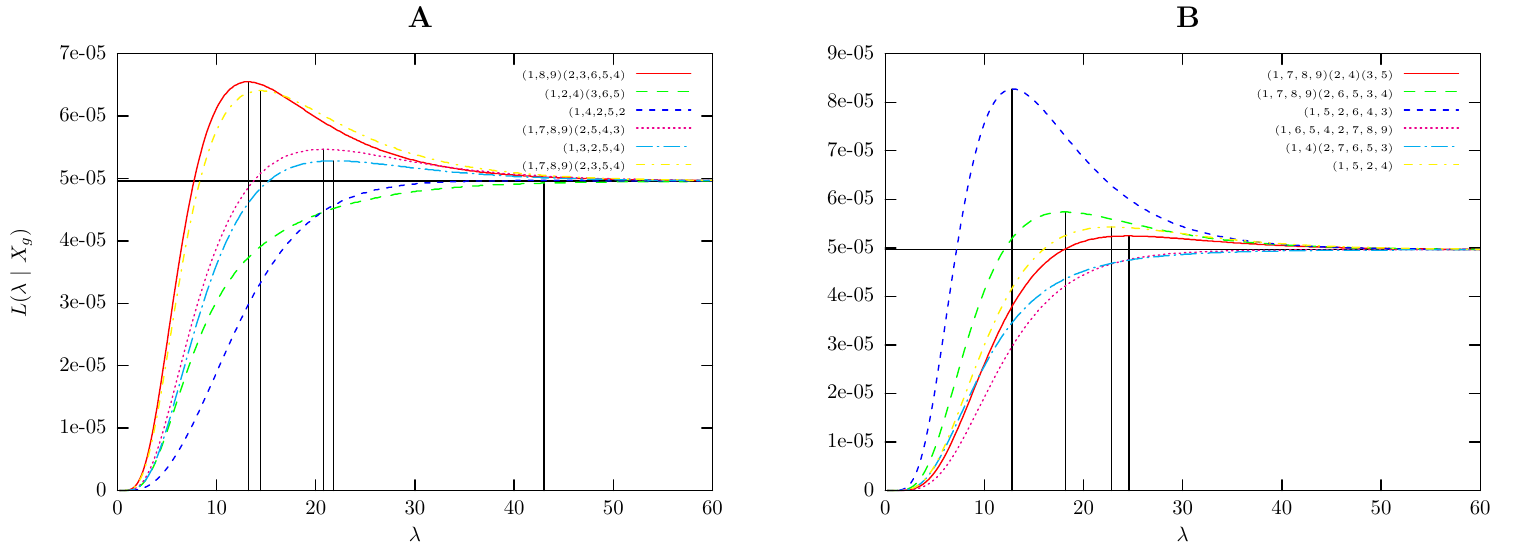}
\caption{Likelihood functions and MLEs where the genome arrangement is viewed as a coset of $\mathcal G$ under $D_n$. 
(A) and (B) show cosets of minimal distance 6 and 7 respectively. The cosets $gD_n$ are labelled by a single representative element.}\label{fig:coset}
\end{figure}

The occasional partial order reversal (with respect to minimal distance) observed for group elements persists for the coset case. This can be seen in Figures~\ref{fig:coset}(A) and~(B), in which MLEs for cosets whose minimal distances are 6 and 7 are shown: some cosets with minimal distance 6 have higher MLE than some cosets with minimal distance 7.   This confirms that minimal distance, even when used on cosets, may be a poor tool for estimating pairwise distance.  

\begin{center}
FIGURE 4 AROUND HERE.
\end{center}

\begin{figure}[!ht]
\centering
\includegraphics[width=\textwidth]{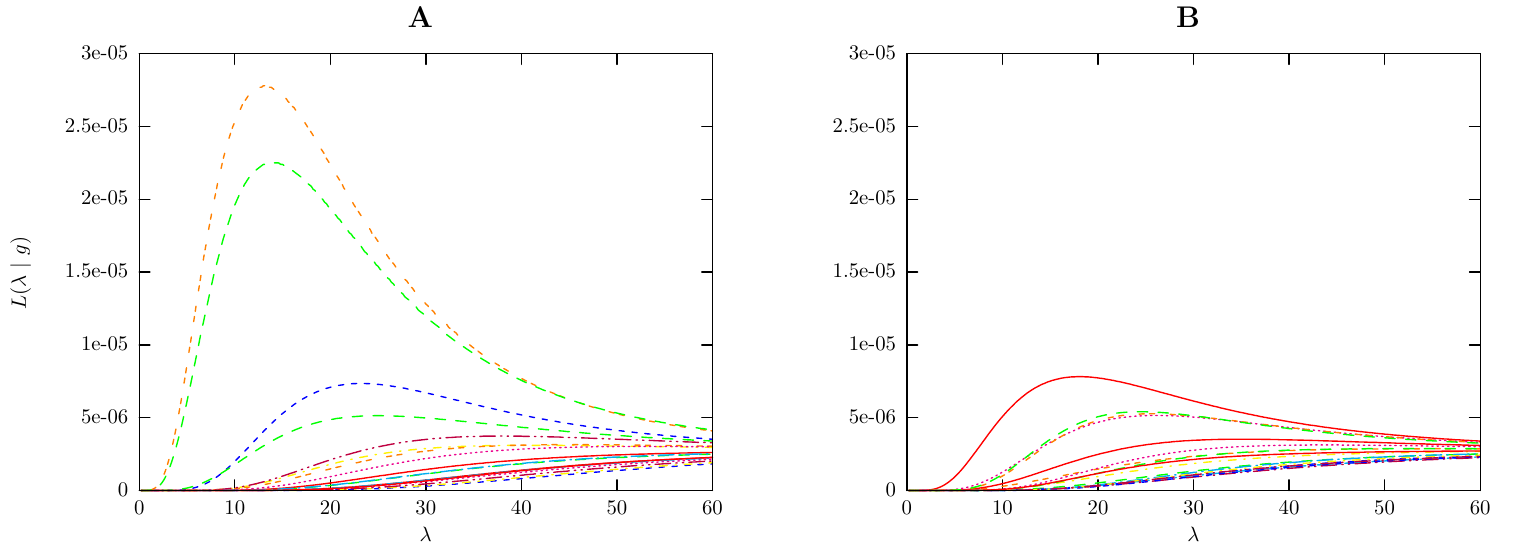}
\caption{Likelihood functions from within two cosets. (A) shows the elements of the coset represented by $(1,2,7)$, while (B) shows the elements of the coset represented by $(1,8,2,9,3,4)$.}\label{fig:samecosets}
\end{figure}

The existence of an MLE for an arrangement is not guaranteed, whether one considers it fixed in space (as a single group element) or free to rotate (as a coset).  Indeed, even if a particular group element gives an MLE, there is no guarantee the coset it resides in also gives an MLE.
Similarly, many cosets with MLEs will contain elements that do not individually possess an MLE.  Figure~\ref{fig:samecosets} shows examples of the likelihood functions for elements within the same coset.  
The two cosets, $(1,2,7)D_9$ and $(1,8,2,9,3,4)D_9$, contain 6 and 5 group elements respectively (out of a total of 18 elements), that individually possess MLEs.  Here, not just the value of the MLE, but also the probability associated with each MLE is important. For a coset to have an MLE, the MLEs of the group elements it contains must be ``strong enough'' to persist through the construction of the coset likelihood function. This illustrates a previous point -- many group elements which themselves possess no MLE, may reside in a coset which \emph{does}.  

\begin{center}
FIGURE 5 AROUND HERE.
\end{center}

For a genome with $n$ regions, each coset  contains $2n$ elements.  The likelihood function for cosets, as in Equation \ref{eqn:cosetlikelihood},  is a sum of the likelihood functions of the elements in the coset.  By accounting for dihedral symmetry, the size of the space reduces by a factor of $2n$: $S_n$ has order $n!$, while $S_n/D_n$ has order $n!/2n$. An exhaustive calculation of $\G = S_9$ shows $\sim 41$\% of group elements possess an MLE. This represents a lower bound on this proportion, as firstly, all terms of the likelihood function are positive (and so a detected maximum will stay a  maximum), and secondly it is possible that new maxima may be found beyond the truncation. By comparison, treating genomes as cosets reveals a lower bound of $\sim 44\%$.

\section{The effect of the use of the coset MLE on phylogeny}

We have seen that the use of a maximum likelihood estimator for evolutionary distance can change the ordering on genome distances.  One would expect this to have a significant effect on phylogenetic inference, and it does.  

\begin{figure}[!ht]
\centering
\includegraphics[width=\textwidth]{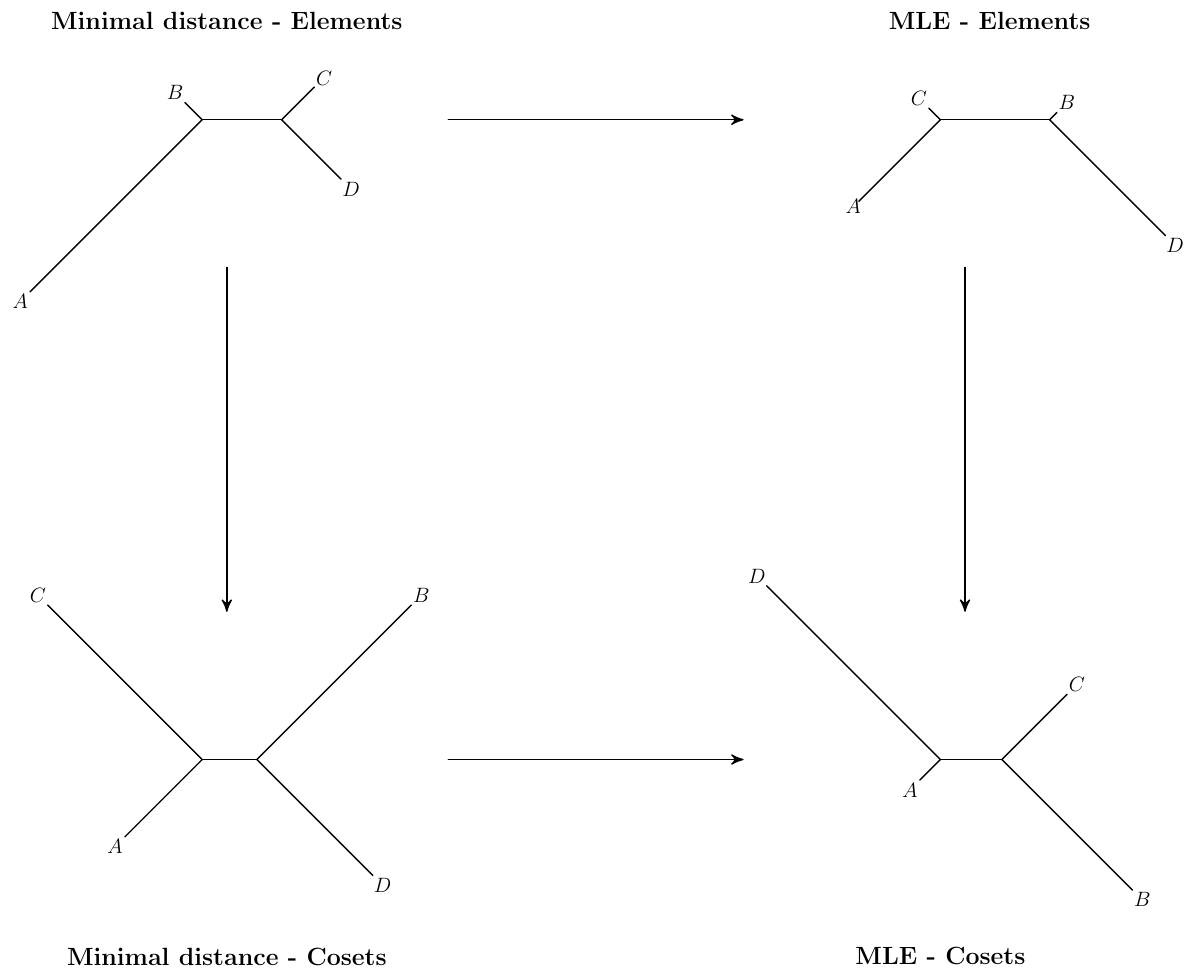}
\caption{Four phylogenies obtained using different distance measures on the same four genomes: $A = ()$, $B = (1,3,2,5,4)$, $C = (1,7,8,9)(2,5,4,3)$, and $D = (1,5,2,6,4,3)$.  
The top left phylogeny uses the fixed frame minimal distance; top right results from using the MLE approach on a fixed frame.  The bottom left represents the coset minimal distance unfixing the frame of reference, and the bottom right results from the MLE approach on cosets. Note that the edge length in the top right diagram between the element $B=(1,3,2,5,4)$ and the bifurcation was negative, an occasional issue with Neighbour Joining.
}
\label{f:phylos}
\end{figure}
The four quartet phylogenies shown in Figure~\ref{f:phylos} differ significantly in their topologies (we use this word in the sense it is used in phylogenetics: the topology of a tree is the arrangement of edges characterizing it, showing the ancestral relationships of the leaves but ignoring edge lengths).  We see $AB|CD$ in the top left (minimal distance on a single group element); $AC|BD$ in the top right and bottom left (MLE on a single group element and minimal distance on a coset); and $AD|BC$ in the bottom right hand corner, our preferred MLE on the coset.  

\begin{figure}
\centering
\includegraphics[width=\textwidth]{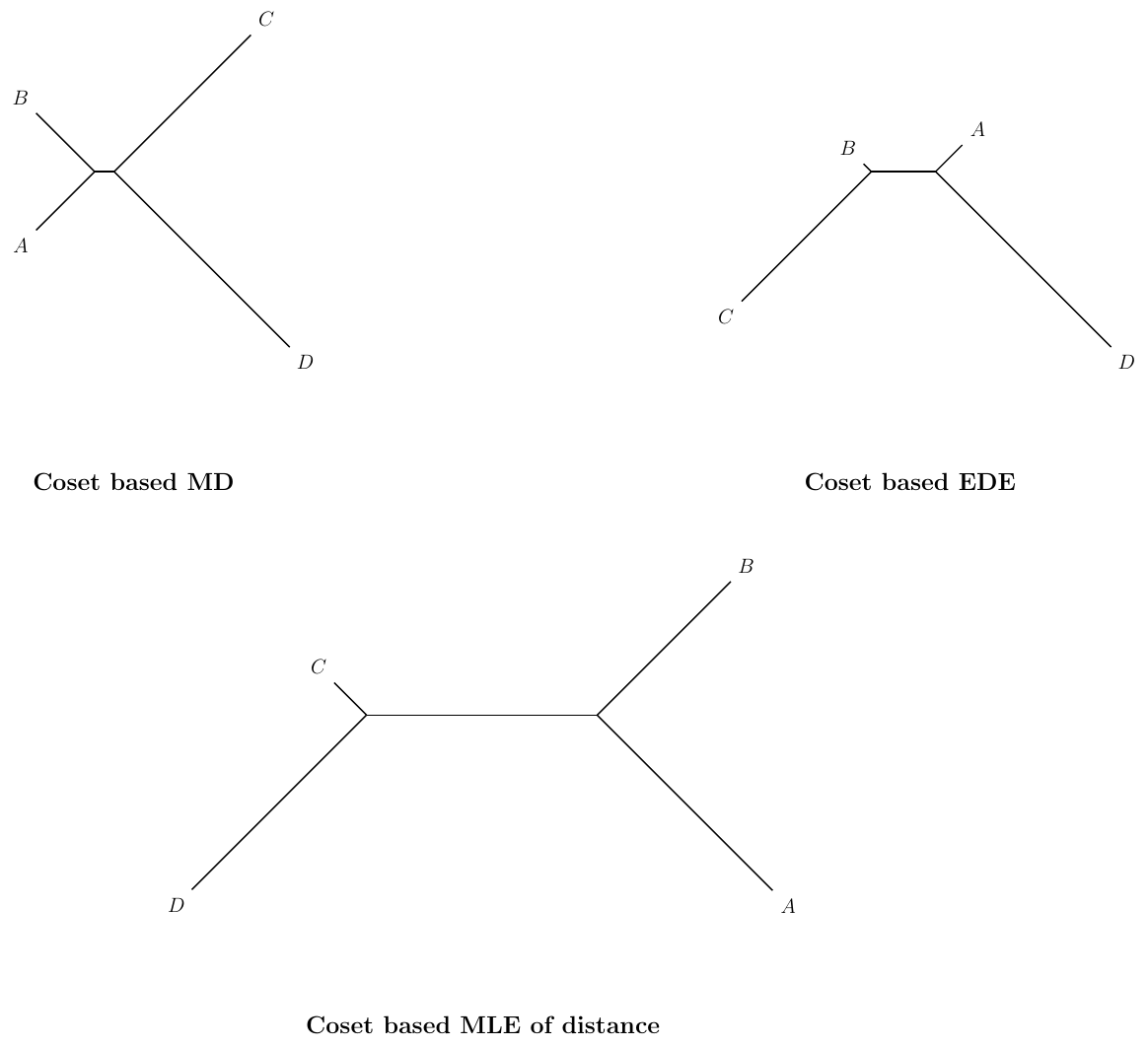}

\caption{Three phylogenies obtained using different coset based distance measures on the same four genomes: $A = ()$, $B = (2,5,4,3)$, $C = (1,3,2,5,4)$, and $D = (1,5,2,6,4,3,7)$. Here we have 3 approaches using the coset methodology: Minimal coset distance;  EDE (based on coset distance); and MLE (coset).  The length of the branch to leaf $B$ in the EDE phylogeny was computed as negative.  (Note: The group elements from the example in Figure~\ref{f:phylos} do not lend themselves to the EDE correction as their minimal distance lies outside the domain of the inverse function.)}
\label{f:phylos.EDE.MLE}
\end{figure}

The coset philosophy can also be paired with other correction methods. The Neighbor Joining phylogenies shown in Figure~\ref{f:phylos.EDE.MLE} are based on distances obtained from three coset methods: the coset minimal distance~\citep{egri2014group}; the \emph{empirically derived estimator} (EDE) of \citet{moret2001new}, using coset minimal distance; and the coset MLE described in this paper.

The coset version of the EDE uses the definition of the minimal distance in coset space. The walk space is sampled and an invertible function is fitted via least squares, to give a one to one relationship between walk length and minimal distance. In the coset sense the only difference from~\cite{moret2001new} is that the minimal \emph{coset} distance (\cite{egri2014group}) is now used instead.

\begin{defn}[Minimal Coset Distance]
Given genome $G$ mapped to group element $g \in \mathcal{G}$ and the corresponding dihedral group $D_n$, the minimal coset distance is defined as 
\begin{align}
d_{cos}(g) = \min\left\{ d_{\min}(gh) ~|~ h \in D_n \right\}
\end{align}
\end{defn}

Using this metric, following the EDE method of Moret \etal, a function is approximated and used to map minimal coset distance to an estimate of true distance. The small sample space allows an exhaustive calculation.

\section{Discussion}\label{sec:disc}

This paper introduces a maximum likelihood estimator for the evolutionary distance between two genomes under a large-scale genome rearrangement model.  One may view this as a correction method for a family of models which can be interpreted using non-abelian groups.  Methods of correcting distances for multiple changes are commonly used, because the use of uncorrected distances can lead to poor inference regarding topology (see \citet{felsenstein2004inferring}).  These corrections for multiple changes are typically implemented in the context of single nucleotide polymorphisms, and these are typically in an environment in which changes at each site are considered to be independent.  The large-scale rearrangements discussed in this paper are different from this in several ways, but the key difference is that rearrangements can affect overlapping regions, and hence interact with each other.  The ``correction'' involved in the context of this paper is to account for evolutionary paths between two genomes that might not be the shortest path.

The MLE approach that we have described requires a selection of generators corresponding to legal rearrangements, and an assumption regarding the probability distribution across these generators. In our examples, we have used a uniform distribution over the set of circular adjacent transpositions for simplicity.  As noted already, the generating set can be readily changed, however it is also possible to change the probability distribution on these generators.  For instance, a natural example might be to model both inversions and translocations, and assign different probabilities to each type of event.  When a non-uniform distribution such as this is selected, the $P(g\mid i)$ factor in Equation~\eqref{eqn:P(GgivenI)} is no longer just a fraction based on walk counts, but can nevertheless still be calculated using similar methods (and with the same complexity). 

The likelihood functions of group elements, and of cosets, do not always have maxima, and in this situation we are not able to give an estimate of evolutionary distance.  This is analogous to the limits of the Jukes-Cantor correction for SNP models of evolution, which are also unable to give a distance when the proportion of nucleotides that have changed exceeds 0.75. An interesting open question would be to characterise those genomes (or group elements) for which the MLE exists, and how this proportion changes for a range of minimal distances.  For instance, while an experiment on genomes of 9 regions under a model of adjacent inversions found slightly fewer than half of the genomes have MLEs (44\%), one would expect that for low minimal distance the proportion would be much higher.  

The key challenges to the MLE approach described here are computational.  The likelihood function for a particular element $g$ consists of two main probabilities: $Pr(i~|~\lambda)$ and $Pr(g~|~i)$. While the first is trivial under our assumptions, the second is computed as the proportion of walks of length $i$ which end at $g$. The current walk count algorithm relies on dynamic programming with complexity exponential in $|\mathcal{S}|$ and memory factorial in $n$. A compromise must be found between truncation and accuracy of the MLE before this can be applied to genomes of realistic numbers of regions (say, greater than 30). The structure provided by the group model gives some hope; Proposition~\ref{prop} is an example of an algebraic property that can be used to significantly reduce computation time. For truncated lookup table construction, this property alone results in a $2n$-fold decrease in computation time.  An approach using group representation theory has recently been developed by some of the authors, and progress in this direction promises to increase the number of regions that the methods in the present paper can handle~\citep{sumner2016representation}.

A second aspect of this work addresses the problem of fixing a frame of reference by taking into account the action of the dihedral group.  This means that we consider arrangements to be equivalent if they can be obtained by physically rotating the genome in three dimensional space.  Algebraically, this involves taking a quotient by the dihedral subgroup and treating each genome as a coset under this quotient. Importantly, this coset approach is not limited to the aforementioned MLE. The ``arrangements are cosets'' view can be taken and applied to other correction methods.  

The coset approach can result in vast differences regardless of underlying metric as shown in small examples. While the MLE may be intractable (without significant algebraic advances in computing or estimating walk counts); the coset approach can be applied with other metrics without changing complexity class. 

\section*{Acknowledgements}
ARF acknowledges support of the Australian Research Council via grants FT100100898 and DP130100248. MMT was supported by Australian Research Council grant FT140100398.

\section*{References}
\bibliographystyle{plainnat}

\appendix

\section{Proof of Proposition~\ref{prop}}
\label{s:proof.of.prop}

\begin{proof}
From Equation~\eqref{eqn:likelihood}, the likelihood functions are of the form $L(\lambda~|~g) = e^{-\lambda}P(\lambda~|~g)$ where $P(\lambda~|~g)$ is a polynomial in $\lambda$. 
Two polynomials are equal if and only if their respective coefficients are equal. Hence we have that 

\begin{align*}
\alpha_i(g_1) = \alpha_i(g_2) \,\, \forall i \geq 0 \implies L(\lambda ~|~ g_1) \equiv L(\lambda ~|~ g_2) \implies \widehat\lambda_{g_1} = \widehat\lambda_{g_2}.
\end{align*}
This reduces the proof to showing that $g_1 \sim_N g_2$ implies $\alpha_i(g_1) = \alpha_i(g_2)$, for all $i \geq 0$. 

Let $g\sim_N h$, so that $h = \pi^{-1} g \pi$ for some $\pi \in N_\mathcal{G}(\mathcal{S})$.  We will show that there is a bijection between the walks of length $i$ to $g$ and to $h$, which will show that $\alpha_i(g)=\alpha_i(h)$ for all $i$.

Let $R_i(g)$ be the set of all length $i$ walks to $g$, so that $|R_i(g)| = \alpha_i(g)$. Take $\gamma_g \in R_i(g)$ to be a walk realised as a concatenation of $i$ generators;
\[ \gamma = s_{k_1}s_{k_2} \ldots s_{k_i}\]
so that
\[ e \gamma = es_{k_1}s_{k_2} \ldots s_{k_i} = g.\]
It follows that 

\begin{align*}
h = \pi^{-1} g \pi 	&= \pi^{-1} e \pi \cdot \pi^{-1} s_{k_1} \pi \cdot \pi^{-1}s_{k_2} \pi \cdot \ldots \cdot \pi^{-1}s_{k_i} \pi. \\
					&= es_{l_1}s_{l_2} \ldots s_{l_i} 
\end{align*}
where $s_{l_n} = \pi^{-1} s_{k_n} \pi$. Because $\pi \in N_{\mathcal G}(\mathcal S)$ and $s_k \in \mathcal{S}$, we have each $s_l \in \mathcal{S}$.  Therefore, $s_{l_1}s_{l_2} \ldots s_{l_i}$ defines a walk to $h$ of length $i$, and so conjugation by $\pi$ is an injective map from $R_i(g) \rightarrow R_i(h)$.  

For surjectivity, each walk $\gamma_h \in R_i(h)$ has pre-image $\pi \gamma \pi^{-1} \in R_i(g)$. Hence	$g_1 \sim_{N} g_2 \implies \alpha_i(g_1) = \alpha_i(g_2)$. 
\end{proof}
\end{document}